# Depositing boron on Cu(111): Borophene or boride?

Xiao-Ji Weng,[1,2] Jie Bai,[1] Jingyu Hou,[1,2] Yi Zhu,[1] Li Wang,[3] Penghui Li,[1] Anmin Nie,[1] Bo Xu,[1*] Xiang-Feng Zhou,[1,2*] and Yongjun Tian[1]

[1]*Center for High Pressure Science, State Key Laboratory of Metastable Materials Science and Technology, School of Science, Yanshan University, Qinhuangdao 066004, China.*

[2]*Key Laboratory of Weak-Light Nonlinear Photonics and School of Physics, Nankai University, Tianjin 300071, China.*

[3]*Vacuum Interconnected Nanotech Workstation, Suzhou Institute of Nano-Tech and Nano-Bionics, Chinese Academy of Sciences, Suzhou 215123, China.*

Large-area single-crystal surface structures were successfully prepared on Cu(111) substrate with boron deposition, which is critical for prospective applications. However, the proposed borophene structures do not match the scanning tunneling microscopy (STM) results very well, while the proposed copper boride is at odds with the traditional knowledge that ordered copper-rich borides normally do not exist due to small difference in electronegativity and large difference in atomic size. To clarify the controversy and elucidate the formation mechanism of the unexpected copper boride, we conducted systematic STM, X-ray photoelectron spectroscopy and angle-resolved photoemission spectroscopy investigations, confirming the synthesis of two-dimensional copper boride rather than borophene on Cu(111) after boron deposition under ultrahigh vacuum. First-principles calculations with defective surface models further indicate that boron atoms tend to react with Cu atoms near terrace edges or defects, which in turn shapes the intermediate structures of copper boride and leads to the formation of stable Cu-B monolayer via large-scale surface reconstruction eventually.

*Corresponding author
bxu@ysu.edu.cn
xfzhou@ysu.edu.cn

## I. INTRODUCTION

In sharp contrast to graphene with a band structure of Dirac semimetal, a two-dimensional (2D) boron allotrope (referred to as borophene) could be a semiconductor, a semimetal or a conductor, depending on the specific polymorph [1-5]. Borophene has aroused enormous research enthusiasm due to the versatile properties such as superior mechanical strength, potential high temperature superconductivity, rare *p*-band antiferromagnetism and intrinsic Dirac fermions, implying important application prospects in related areas [6-10]. Bulk boron allotropes usually are composed of $B_{12}$ icosahedra and interstitial boron atoms with a three-dimensional strong covalent network structure. It is apparent that borophene may not be directly exfoliated from a parent bulk phase but may be achieved by depositing boron or other precursors on suitable substrate. Note that boron can hardly form compounds with group IB (Cu, Ag, Au) due to the small difference in electronegativity as well as large difference in atom size between boron and these metal elements. Therefore, group IB metals are preferentially selected as the substrates for borophene preparations.

Experimentally, the first monolayer borophene was synthesized in 2015 by depositing boron on Ag(111) substrate with molecular beam epitaxy (MBE) under ultrahigh vacuum [6,11]. The following years have witnessed a series of borophene syntheses on different metal substrates, such as Ag(110) [12], Ag(100) [13], Au(111) [14], Ir(111) [15], Cu(111) [16-19] and Cu(100) [20]. Among these substrates, Cu is of particular interest due to its low cost and weak interaction with the supported 2D materials, which is important for scalable fabrication and different transfer processes to technologically relevant substrates [18]. Since 2019, several independent works have synthesized large-area (up to millimeters) single-crystal surface structures on Cu(111) substrate with B deposition [16-19,21]. Meanwhile, several 2D copper borides were theoretically predicted with interesting electronic and catalytic properties [22,23]. While borophenes were claimed to form on Cu(111) in most experimental works [16-19], an unexpected copper boride was argued to form with similar growth conditions [21]. The first successful synthesis of borophene was reported by depositing boron on Cu(111) at *ca*. 770 K with MBE [16], where a $\chi_3$-like borophene structure was proposed. Later, this borophene structure was produced on Cu(111) with a straight-forward CVD approach using diborane precursor [18]. Most recently, a distinct $\beta_{12}$-like borophene structure was proposed for the MBE-produced monolayer borophene on Cu(111) above 600 K, where a large-size bilayer borophene formed with increasing boron coverage [19]. In comparison, we also grew similar surface structure on Cu(111) at *ca*. 740 K with MBE, and attributed it to a 2D $Cu_8B_{14}$ structure though [21]. Whether this surface structure is a $\chi_3$-like borophene, $\beta_{12}$-like borophene, or an unexpected copper boride remains controversial [16,19,21].

In this work, we compared scanning tunneling microscope (STM) observations from different works along with the respective structural models and simulated STM images. Surface-sensitive characterizations such as X-ray photoelectron spectroscopy (XPS) and angle-resolved photoemission spectroscopy (ARPES) measurements were conducted to screen out the possible phase of borophene or copper boride. Furthermore, first-principles calculations were performed to elucidate the interaction between defective Cu(111) surface and deposited boron atoms. Our systematic investigation unambiguously reveals the formation of copper boride on Cu(111) after B deposition.

## II. RESULTS AND DISCUSSION

STM characterizations of boron deposited structures on Cu(111) from independent works were shown in Fig. 1 along with the proposed structures [16,19,21]. The STM images [Figs. 1(a)–1(c)] show distinct zigzag patterns. Nonetheless, similar unit cells with lattice constants of $a \approx 21.8$ Å, $b \approx 16.0$ Å and $\alpha \approx 70°$ were identified for these surface structures, corresponding to a $\sqrt{73} \times \sqrt{39}$ superstructure on Cu(111). Considering the similar growth conditions, it is likely that these seemingly different STM images are from a same surface structure, with the differences induced by different scanning conditions such as temperature, tip conditions, bias voltage and tunneling current. To account for the STM observed surface structure, several 2D structural models were proposed based on different considerations, *i.e.*, two borophene structures with distinctly mixed triangular and hexagonal motifs [Figs. 1(d) and 1(f)] and one copper boride [$Cu_8B_{14}$, Fig. 1(e)]. The simulated STM images based on the structural models are shown in Figs. 1(g)–1(i). The first borophene structure [Fig. 1(d)] is a $\chi$-type borophene [16]. In the unit cell, there are 52 boron atoms with coordination number (CN) of either 4 or 5, denoted as $\chi$-52 borophene. In this model, the hexagonal vacancies form zigzag lines with an intersection angle of 88°, slightly larger than 84° from STM measurement [Figs. 1(a) and 1(d)]. Particularly, the intersections of the unit cell diagonals show obviously different contrast in experimental and simulated STM images [Figs. 1(a) and 1(g)], indicating an inconsistency between the surface structure and the proposed model. The second borophene structure [Fig. 1(f)] is a β-type one (CN = 4, 5 or 6) with 62 boron atoms in the unit cell [19], denoted as β-62 borophene. In this model, the zigzag lines running through 6-coordinated B atoms show an intersection angle of 120°, substantially deviated from the experimental value of *ca*. 110° [Figs. 1(c) and 1(f)]. Given the fact that the distance between two adjacent bright spots [about 2.5 Å, Fig. 1(b)] is very close to the nearest neighbor distance of Cu atoms, we proposed the formation of a copper-rich boride, $Cu_8B_{14}$, on Cu(111) substrate instead of borophene [Fig. 1(e)], which is thermodynamically more stable than χ-52 borophene [21]. The simulated STM image of $Cu_8B_{14}$ matches the experimental image nicely with the zigzag pattern determined by Cu atoms in copper boride showing an intersection angle of 94° [Figs. 1(b) and 1(h)]. Interestingly, the zigzag lines through B atoms show an intersection angle of *ca*. 84° [see the blue lines in Fig. 1(e)], very close to the experimental value observed in Fig. 1(a). Overall, our copper boride model shows better agreement with STM observation. To further eliminate the controversy in the structure, other experimental characterization techniques in addition to STM were conducted.

XPS is a powerful technique to identify the chemical state and stoichiometry of the surface structure. Previously, ex-situ XPS spectra showed that about 80% of "borophene" on Cu(111) could be oxidized after one hour of exposure to air [16]. To exclude the possible oxidation, we conducted in situ angle-resolved XPS measurements of Cu(111) before and after boron deposition. As shown in Fig. 2(a), the clean Cu(111) showed Cu $2p_{3/2}$ core level peaked at 932.7 eV (consistent with the ISO standard Cu metal line at 932.63 eV), without any noticeable change at various emission angles (the angle between emission electron and sample surface). In comparison, Cu $2p_{3/2}$ core level after boron deposition shifted slightly to higher binding energy, indicating a charge transfer from copper. More interestingly, Cu $2p_{3/2}$ peak substantially broadened (to the higher binding energy direction) with decreasing emission angle, suggesting an enrichment of copper species on the surface that is different from the bulk Cu and more sensitively detected under grazing emission angle. Meanwhile, B 1s core level peaked at 187.8 eV did not show noticeable broadening at various emission angles, indicating a simple chemical state of boron atoms. Therefore, the surface structure formed on Cu(111) substrate after boron deposition most likely contains Cu atoms, which interact more strongly with B atoms and become more pronounced in XPS spectra taken at smaller emission angle. To quantify the ratio between boron and copper in the topmost surface, Cu $2p_{3/2}$ and B 1s core levels were fitted with two peaks and one peak, respectively. The two fitting peaks for Cu $2p_{3/2}$ denoted as P1 and P2 correspond to bulk Cu and surface Cu, respectively [Fig. 2(b)]. It is obvious that the surface component P2 significantly increased as the emission angle decreased from 75° to 10° due to the surface sensitivity of XPS. A binding energy difference of *ca*. 0.6 eV was revealed between bulk Cu and surface Cu. As a comparison, the binding energy difference is 0.9 eV for $2p_{3/2}$ between Cu and CuO [24]. Considering the relative sensitivity factor (RSF) of Cu $2p_{3/2}$ and B 1s, the atomic ratio between surface Cu and B atoms, defined as $(I_{P2}/RSF_{Cu-2p_{3/2}})/(I_B/RSF_{B-1s})$, ranged from 0.51 to 0.61 for different emission angles, in a good agreement with the ratio of 0.57 (8/14) for $Cu_8B_{14}$ [21]. The evidence from XPS measurements clearly indicates that, on Cu(111) substrate after boron deposition, a new Cu species exists and strongly interacts with boron, consistent with the proposed $Cu_8B_{14}$ surface structure [21].

ARPES measurement was conducted on the surface copper boride, *i.e.*, $Cu_8B_{14}$, to further explore the electronic property. The iso-energy ARPES mapping was shown at different binding energy, where a stronger difference in contrast was observed at binding energy lower than -0.6 eV [see Supplemental Material (SM) for more details [25]. A three-fold rotational and reflection symmetry is revealed from the ARPES maps [Fig. 3(a)], which is obviously related to the Cu(111) surface of three-fold symmetry and six different overlayer domains (two mirror-symmetric ones for each of three high-symmetry directions of Cu surface, as confirmed by the low energy electron diffraction measurement [21]). Figure 3(c) exhibits the ARPES spectra along Cu $[1\bar{1}0]$ direction [see cut1 in Fig. 3(b)], showing a metallic character apparently distinct from that of clean Cu(111) [26]. Specifically, the valence bands can be classified into two groups: The M-shaped bands (yellow dashed lines) around Γ-point and a series of U-shaped bands (white dashed lines) located at the momentum $k$ of 0.2–0.6 Å$^{-1}$. To understand the origin of these bands, a twelve-layer-thick slab model including ten (111) layers of Cu and two surface layers of symmetrically distributed copper boride was constructed for band structure calculation [Fig. 3(d)]. Due to time-reversal symmetry, three non-equivalent orientations of $Cu_8B_{14}$ were considered in the calculation. The calculated band structure along cut1 was projected to different atomic layers to identify the respective contribution from bulk or surface [Figs. 3(e) and 3(f)]. Clearly, the M-shaped bands are dominantly contributed by layer3 atoms [Fig. 3(e)], while the U-shaped bands are mainly contributed by layer1 atoms, *i.e.*, the surface layer of copper boride [Fig. 3(f)].

Up to now, STM, XPS and ARPES characterizations have evidenced the formation of copper boride on Cu(111) after boron deposition. However, the formation mechanism of this binary surface structure of Cu and B, which traditionally are considered as immiscible elements, remains unclear. Previously, the lowest-energy configurations of boron clusters on ideal Cu(111) surface were investigated theoretically [27], where borophene formation was preferred due to increasing cluster size and low diffusion barrier. In experiment, real-time monitoring of boron deposition on Cu(111) with low energy electron microscopy revealed preferential nucleation of small islands on the lower terraces of step edges [16], implying the possible involvement of Cu atoms from the step edges. We thus proposed a defective surface model, which is more in line with the realistic condition, to qualitatively probe the growth process [Fig. 4(a)]. As described in Methods section, tens of thousands of surface structures were produced and relaxed via *ab initio* evolutionary searches. Four of the most thermodynamically stable structures are presented in Figs. 4(b)–4(e), and others in SM [25]. From the analysis, we can draw the following conclusions. First, almost all the stable structures are formed by reacted B and Cu atoms near terrace edges or defects. Second, some B atoms may penetrate into Cu substrate, such as in the cases with $N_B$ = 1, 4, 5 and 7. Third, some Cu atoms from the second layer can be pulled out due to B adsorption, such as $N_B$ = 2, 7, 8 and 9. Fourth, substructures of boron chains or nanobelts tend to form near terrace edges or defects, such as $N_B$ = 3, 4, 6, 7 and 9. Fifth, small fragments of 2D copper boride appear due to the surface reconstruction, such as $N_B$ = 5, 7, 8 and 9. In short, copper boride surface structures dominantly form on defective Cu(111) substrate after B deposition. To evaluate the thermodynamic stability of the above mentioned nine surface structures in comparison with $Cu_8B_{14}$, the formation energy was calculated, indicating that clean Cu(111) and $Cu_8B_{14}$ are thermodynamically stable phases, which is consistent with experimental observations and early prediction [21]. Other Cu-B surface structures are metastable with respect to Cu and $Cu_8B_{14}$, implying possible existence as intermediate phases during growth of copper boride.

In summary, systematic experimental characterizations combined with first-principles calculation clarify the controversy of the surface structure formed on Cu(111) after B deposition and elucidate the formation mechanism of the unexpected copper boride. The large-area single-crystal surface structure on Cu(111) should be attributed to the boron deposition induced surface reaction, *i.e.*, the formation of $Cu_8B_{14}$ (a 2D copper boride). Our work strongly suggests that caution must be exercised when proposing structure models on limited information. On the other hand, the discovery and structure determination of this 2D copper boride (as an emergent material) are of fundamental importance, and shed light on preparation of other planar systems, such as monolayer $Cu_2B_2$ (promising for CO reduction to $C_2$ products [23]) and $Cu_7B_{15}$ (an intrinsic nodal line semimetal [22]).

ACKNOWLEDGMENTS

This work was supported by the National Natural Science Foundation of China (Grant Nos. 52025026 and 11874224), the National Key R&D Program of China (2018YFA0305900) and the Natural Science Foundation of Hebei Province of China (E2022203109). The authors are grateful for the technical support from Nano-X from Suzhou Institute of Nano-Tech and Nano-Bionics, Chinese Academy of Sciences (SINANO).

**APPENDIX: METHODS**

**1. Sample preparation and measurement**

Cu(111) substrate was treated with repeated cycles of argon ion sputtering and sequential annealing for a clean and flat surface, and then transferred to MBE chamber with a base pressure of $2\times10^{-10}$ mbar for boron deposition. Pure boron (99.9999%, Alfa Aesar) was evaporated with a high temperature effusion cell or electron-beam evaporator. After boron deposition, the sample was transferred to the conjoint characterization chamber under ultrahigh vacuum for further STM, XPS and ARPES measurements. Large-area single-crystal phase was achieved on Cu(111) surface at *ca.* 700 K, as confirmed by the room-temperature STM (Aarhus VT-STM) measurement. Angle resolved XPS measurement was performed on a PHI 5000 VersaProbe III system with monochromatic Al $K_\alpha$ radiation. Cu $2p_{3/2}$ and B $1s$ core level peaks were fitted with an asymmetric Gaussian–Lorentzian mixture function. ARPES measurement was performed on a DA30L ARPES system with monochromatic He I (21.2 eV, energy resolution < 1.8 meV, angle resolution < 0.1°).

**2. Calculation**

The *ab initio* DFT calculations were conducted with the projector augmented wave method as implemented in the VASP code [28,29]. The exchange correlation energy was treated within the generalized gradient approximation using the functional of Perdew, Burke and Ernzerhof [30]. Plane-wave cutoff energy of 450 eV, Γ-centered k-point grid with a resolution of $2\pi \times 0.04$ Å$^{-1}$, electronic self-consistency cycle convergence criterion of $10^{-5}$ eV and force criterion of $10^{-2}$ eV/Å were used for the structure relaxation. The STM images were simulated according to the Tersoff–Hamann theory and visualized in the constant current mode with p4VASP program [31]. A slightly distorted $2\times\sqrt{37}$ Cu(111) supercell was adopted for ARPES simulation, where $Cu_8B_{14}$ monolayers are symmetrically placed on both sides of a ten-layer-thick Cu(111) slab to avoid the Cu(111) surface state. Layer-projection and *k*-projection of band structure (band unfolding) were carried out with KPROJ program [32]. To qualitatively uncover the growth mechanism involved with surface defects or terraces, a modified three-layer-thick $2\times3\sqrt{3}$ Cu(111) slab was constructed for surface structure search with the evolutionary algorithm USPEX [33-35]. To simulate terrace or defect on the surface, half of topmost Cu atoms were deleted intentionally in this model. Different numbers of boron atoms ($N_B$, up to 9 per surface unit cell) were deposited on the model surface within a thickness of 2.5 Å, and allowed to react with the topmost and defect Cu atoms during the relaxation. Meanwhile, the bottom Cu layer was fixed, and the total energy was taken as a criterion for generation of new structures from random generators, heredity, and mutations. The formation energy for all surface structures was calculated as

$$E_f = \left(E_{tot} - E_{sub} - N_{Cu}\mu_{Cu} - N_B\mu_B\right)/A,$$

where $E_{tot}$ and $E_{sub}$ are the total energy of the system and the two-layer-thick Cu(111) slab, respectively, $N_{Cu}$ is the number of Cu atoms in the topmost layer (*i.e.*, $N_{Cu}$ = 6 for all the defective configurations and $N_{Cu}$ = 32 for $Cu_8B_{14}$ over a $\sqrt{73}\times\sqrt{39}$ supercell on Cu(111) surface), $N_B$ is the number of B atoms, $\mu_{Cu}$ is the chemical potential of copper equal to the energy per atom for face-centered cubic Cu, $\mu_B$ is the chemical potential of boron and $A$ is the area of surface structure.

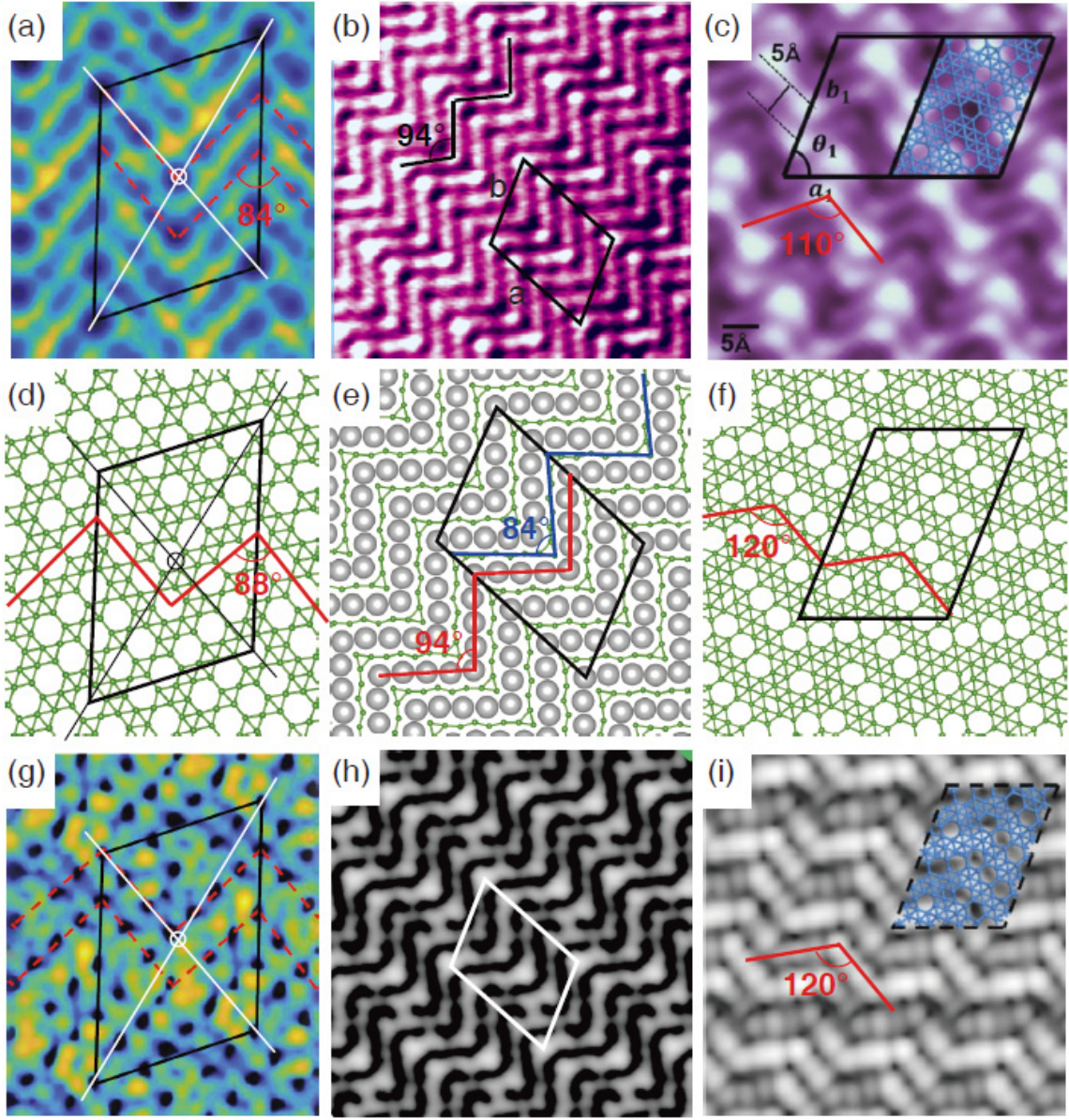

FIG. 1. STM images of different samples after depositing boron on Cu(111). (a) STM image from Ref. 16 ($V_t$ = +20 mV, $I_t$ = 50 pA, 5K, CO-functionalized Pt/Ir tip). Copyright 2019 Nature. (b) STM image from our experiment ($V_t$ = +110 mV, $I_t$ = 80 pA, room temperature, W tip). (c) STM image from Ref. 19 ($V_t$ = -100 mV, $I_t$ = 210 pA, 77 K, W tip). Copyright 2022 Nature. (d)–(f) Three proposed monolayer structures, χ-52 borophene, $Cu_8B_{14}$, and β-62 borophene, respectively. Green and gray spheres indicate boron and copper atoms, respectively. (g) The simulated STM image from Ref. 16. Copyright 2019 Nature. (h) The simulated STM image ($Vt$ = +1.0 V) based on $Cu_8B_{14}$ structure of (e). (i), The simulated STM image from Ref. 19. Copyright 2022 Nature. The labelled marks highlight the intersections of the unit cell diagonals and the zigzag patterns with specific intersection angles.

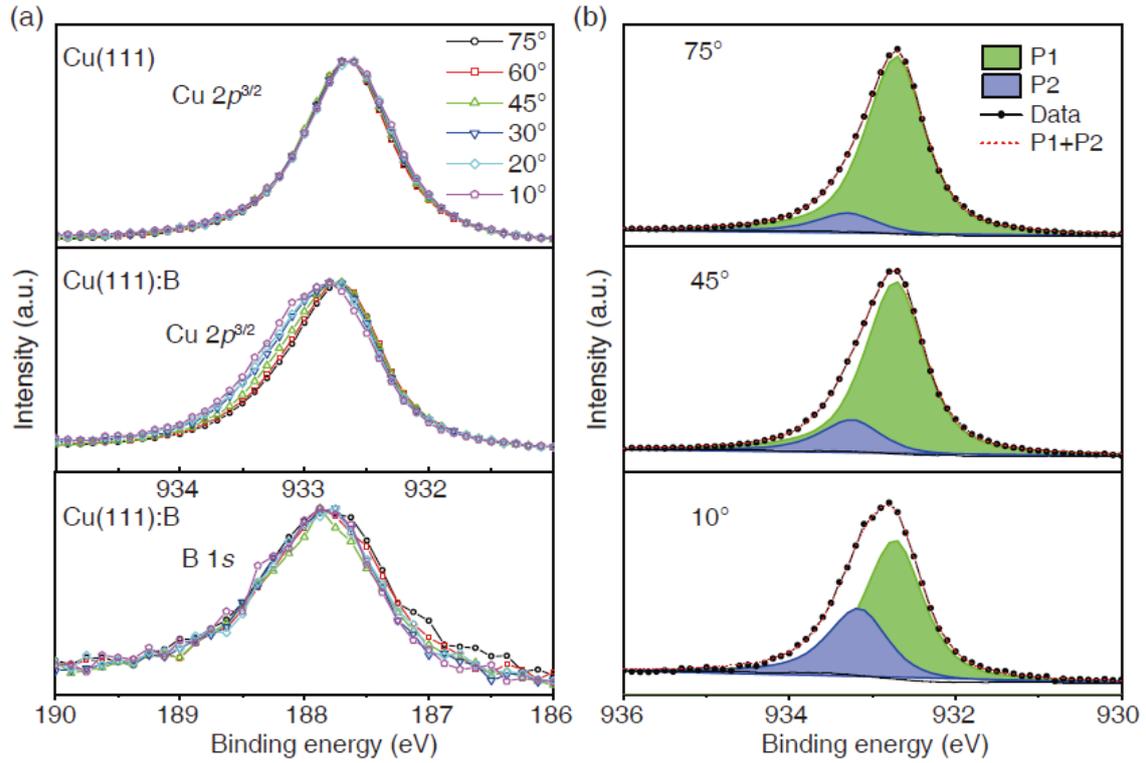

FIG. 2. Angle-resolved XPS spectra of Cu-B system and clean Cu. (a) Cu $2p_{3/2}$ and B $1s$ core level peaks at various emission angles (Upper panel, Cu $2p_{3/2}$ from clean Cu(111) surface; middle panel, Cu $2p_{3/2}$ after boron deposition; bottom panel, B $1s$). The peak intensities were normalized. (b) The peak fitting for Cu $2p_{3/2}$ core level taken at different emission angles (upper panel, 75°; middle panel, 45°; bottom panel, 10°).

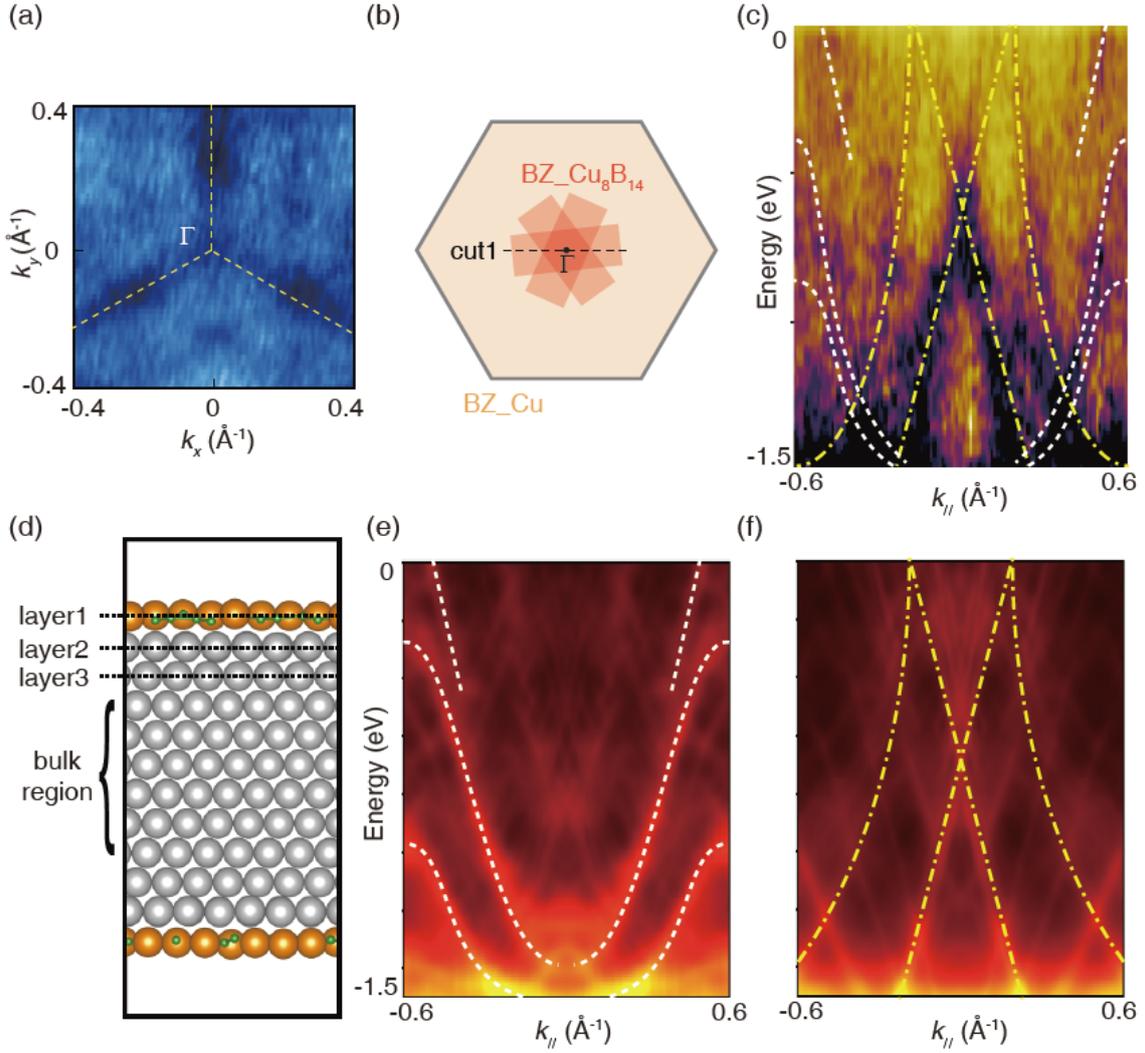

FIG. 3. ARPES spectra of copper boride on Cu(111). (a) ARPES intensity map near Γ taken at 0.6 eV binding energy. (b) The Brillouin zone (BZ) of and pristine Cu(111) and overlapped $Cu_8B_{14}$. (c) ARPES intensity map taken through Γ and perpendicular to $k_x$, *i.e.*, cut1 direction ($k_{//}$) in (b). (d) The structural model used for ARPES simulation, in which the $Cu_8B_{14}$ monolayers are symmetrically placed on both sides of ten-layer-thick Cu(111) slab. The topmost $Cu_8B_{14}$ monolayer and the second highest Cu(111) layer were denoted as layer1 and layer3, respectively. (e) and (f) The calculated band structures along cut1 contributed from layer 1 and layer 3, respectively.

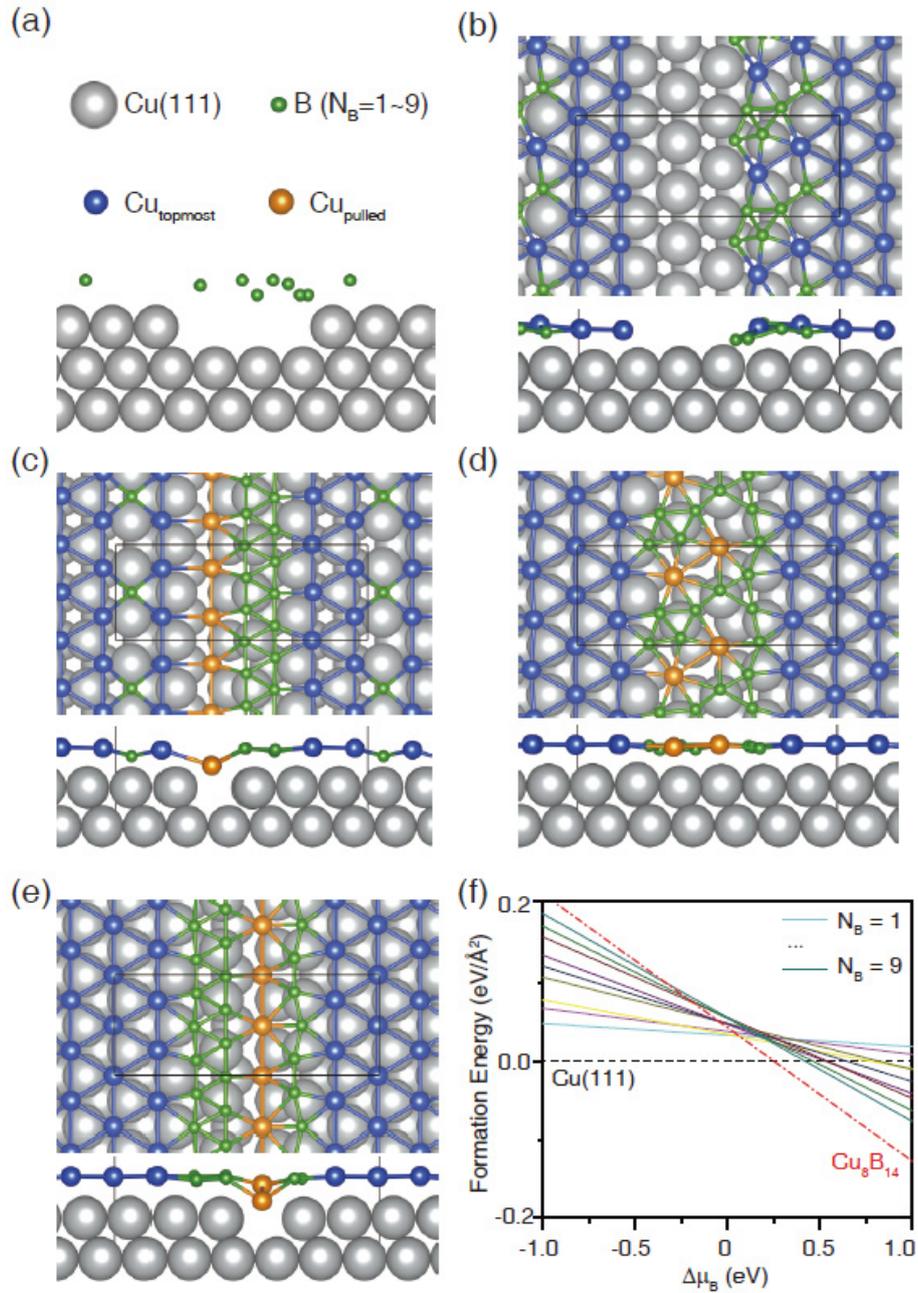

FIG. 4. Surface structure prediction for depositing boron on defective Cu(111). (a) Schematic diagram of randomly distributed boron atoms over defective Cu(111). (b)–(e) Structures with the lowest energy from the structure search. The numbers of B atoms per unit cell, $N_B$, equal to 5, 7, 8 and 9, respectively. Cu atoms belong to bulk Cu and the topmost Cu(111) layer are colored in gray and blue, respectively, other Cu atoms (pulled up after boron adsorption) are colored in yellow. (f) Formation energy of various surface structures (with respect to pristine Cu(111) surface) as a function of boron chemical potential. The energy of bulk α-$B_{12}$ is set as the origin.

# Supplementary Materials for

# Depositing boron on Cu(111): Borophene or boride?

Xiao-Ji Weng,[1,2] Jie Bai,[1] Jingyu Hou,[1,2] Yi Zhu,[1] Li Wang,[3] Penghui Li,[1] Anmin Nie,[1] Bo Xu,[1*] Xiang-Feng Zhou,[1,2*] and Yongjun Tian[1]

[1]*Center for High Pressure Science, State Key Laboratory of Metastable Materials Science and Technology, School of Science, Yanshan University, Qinhuangdao 066004, China.*

[2]*Key Laboratory of Weak-Light Nonlinear Photonics and School of Physics, Nankai University, Tianjin 300071, China.*

[3]*Vacuum Interconnected Nanotech Workstation, Suzhou Institute of Nano-Tech and Nano-Bionics, Chinese Academy of Sciences, Suzhou 215123, China.*

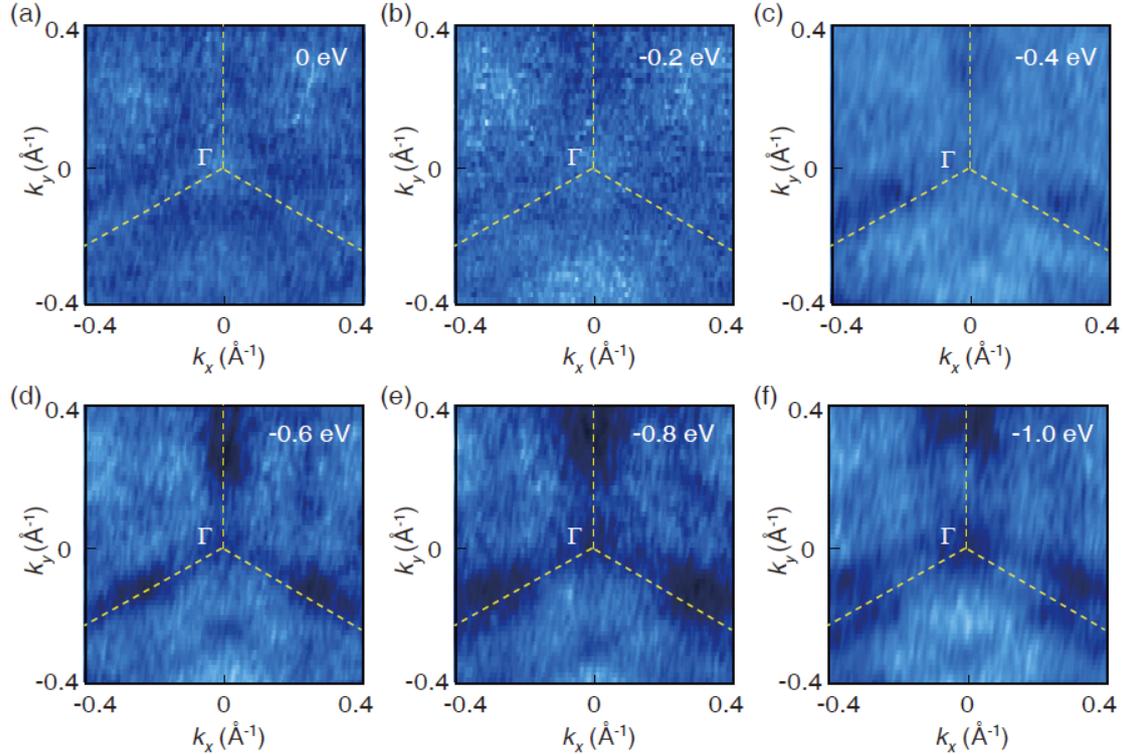

**FIG. S1.** ARPES intensity map near Γ taken at energy of (a) 0.0, (b) -0.2, (c) -0.4, (d) -0.6, (e) -0.8, (f) -1.0 eV ($E_{Fermi} = 0$).

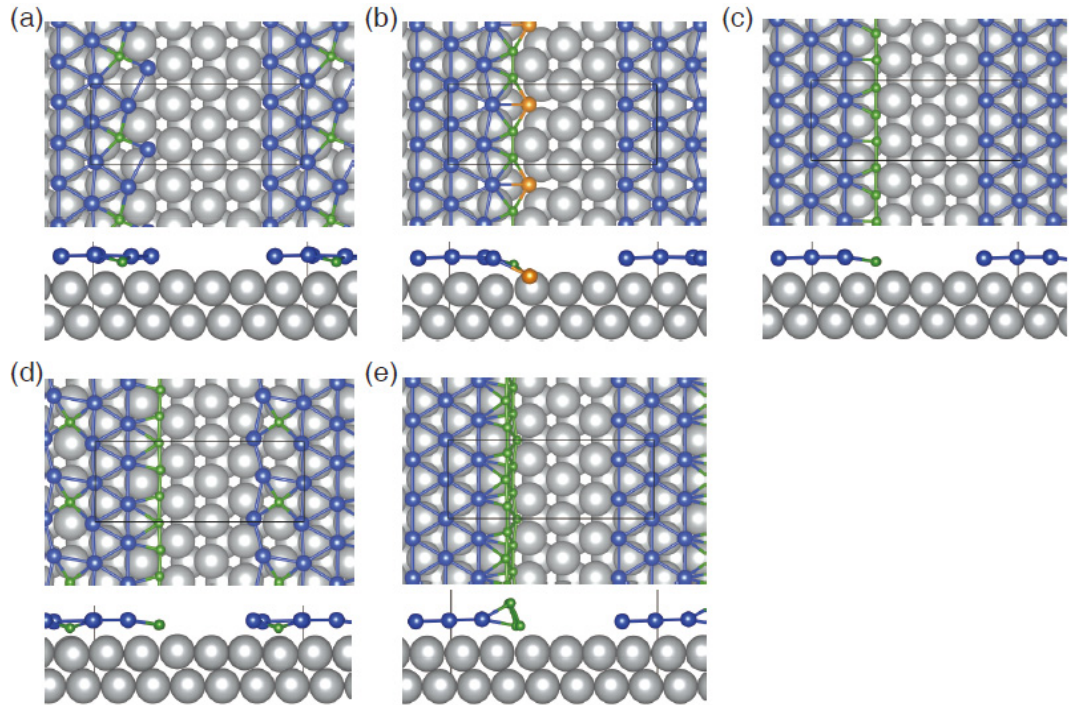

**FIG. S2. The configuration with the lowest energy from the structure searches for boron on defective Cu(111).** $N_B$ = (a) 1, (b) 2, (c) 3, (d) 4, (e) 6, respectively. The Cu atoms belong to bulk Cu(111) and the topmost Cu(111) layer were colored in gray and blue respectively, while other Cu atoms pulled up by boron adsorption were colored in yellow.